\documentstyle[aps,preprint]{revtex}
\tighten
 
\begin{document}
    \draft
    \title{
    Formula for the widths of the plateaus of the quantum Hall effect
     \\
    }
    \author{ Atanas Groshev$^{a,b}$ }

    \address{%
    $^a)$ Institut f\"ur Theoretische Festk\"orperphysik,
    Universit\"at Karlsruhe, \\
    76128 Karlsruhe,  Federal Republic of Germany\\
    $^b)$ Department of Theoretical Physics,
University of Sofia, 5 J. Boucher Blvd., \\
1126 Sofia, Bulgaria.
    }

\maketitle

    \begin{abstract}

We derive an empirical formula for the width of quantum Hall effect
plateaus which is free of adjustable parameters.
It describes  the integer, fractional and $\nu = 0$
(Wigner insulator) quantum Hall effect in single heterojunctions.
The temperature scale of the existence
of these three phenomena is the same as the melting
temperature of a classical Wigner crystal.
We conclude that the basic
assumption of the current theory of QHE that the plateau width
is determined by the disorder is highly improbable.

\end{abstract}


\narrowtext

Three phenomena of completely different nature are believed
to exist in a 2-dimensional electron system (2DES) in high
magnetic fields at low temperatures
\cite{pra1}. The integer quantum Hall
effect is thought to be describable as  single particle localization
(by the disorder)
of a noninteracting electron gas \cite{pra1}. The fractional Hall effect
is taught to be due to quasiparticle localization
(by the disorder) on top of a many-particle liquid state \cite{pra1}.
A Wigner crystal (many-body effect) exists at low filling factors
\cite{sha1}.
The essence of the quantum Hall effect (QHE) is the existence of
plateaus of {\it finite} width in
the Hall resistance \cite{kli1}.
In spite of substantial theoretical efforts devoted to this subject no
successful
formula has yet been derived for the
width of these plateaus. However, since it is believed that the plateau
width is determined by localization in the disorder potential,
one would expect it to be sample dependent and moreover to
decrease and disappear in the low disorder limit.

Here we discuss experimental facts which contradict
this assumption. On the basis of the symmetries of the experimental
data we derive an empirical formula
for the width of the plateaus of the quantum Hall effect
(integer and fractional) as well as for the width of the filling factor
region of the Wigner solid.We also show that the temperature scales
of the integer QHE, fractional QHE and the Wigner solid are
the same and close to the classical melting temperature of the
Wigner crystal.

The Landau level filling factor of the 2DES of density $n_s$ in a
magnetic field $B$ is given by
$\nu = n_s h c/eB$. In the absence of the quantum Hall effect
(for example at high temperatures)
the Hall conductivity is given by the classical value
$\sigma_{xy} = e n c/B = \nu e^2/h$. The quantum Hall effect
plateaus appear at integer and fractional values of
$\nu$.
 The most important factors influencing the width of plateaus are the
temperature $T$ and  the disorder \cite{rem1}.
  We will be interested in the limit of very low
temperatures and disorder. We will quantify the meaning of ``very low''
below.

 We start with several observations concerning the shape
 of the dependence $\sigma_{xy}(\nu)$ in this limit.
 They are based on visual examination of the available data
 and should be considered as empirical rules.\\
1. The transitions between adjacent plateaus are sharp steps.\\
2. The  plateaus extend symmetrically  on both sides
   of  the classical line $\sigma_{xy}(\nu) = \nu e^2/h$
   (see Fig.~\ref{platofg1}).
   We denote the half-width of the plateau at the filling
   factor $\nu$ by $\Delta\nu_{\nu}$. \\
3. The following symmetry relations are fulfilled
   \begin{equation}
   \Delta\nu_{1\pm\nu} = \Delta\nu_{\nu}
   \label{s}
   \end{equation}
4. The plateaus are grouped in sequences
   as follows.\\
   A main sequence converging towards 1/2:
   $$ \nu = p/(2p+1),\ \ \ p = 0,1,2\ldots $$
   and its symmetry partners according to (\ref{s}).
   This is the only sequence in 2D hole systems.\\
   Two additional secondary sequences converging towards 1/4
   exist in 2D electron systems:
   $$ \nu = p/(4p\pm 1),\ \ \ p = 0,1,2\ldots $$
   and their symmetry partners according to (\ref{s}).\\
   The positions of these plateaus is given by
   the single formula $\nu_{m,p} = p/(2 m p\pm 1)$ where $m = 1$
   corresponds to the main sequence and $m = 2$ to the secondary
   sequences.\\

We would like to stress that the property 4. has been discussed
earlier in the literature  \cite{cha1,wil1,du1,jai1,kiv1,hal1}.
The symmetry $\nu \rightarrow
1\pm \nu$ has also been discussed earlier \cite{kiv1,hal1} although
in the present form (\ref{s}) we give it for the first time.
To our knowledge the properties 1. and 2.
have not been discussed earlier,
although we consider them apparent from the experimental
data (see for example \cite{dav1,wil1,saj1}).

Using these facts we find a formula for the widths
of the plateaus.
Indeed, let $\nu_{m,p}$ and $\nu_{m,p+1}$ be two adjacent plateaus within
a particular sequence.
As a consequence of 1. and 2., the following recursive equation holds
for the plateau widths (see also Fig.~\ref{platofg1}):
\begin{equation}
\Delta\nu_{\nu_{m,p}} + \Delta\nu_{\nu_{m,p+1}} = |\nu_{m,p+1} - \nu_{m,p}|
\label{equ}
\end{equation}
The solution of this recursive equation
is unique for each of the sequences and is given by the formula:
\begin{equation}
\Delta\nu_{p/(2mp \pm 1)} = \frac{1}{8m^2} \left[
2 \psi\left({p\over 2} + {1\over 2} \pm {1\over 4m}\right) -
\psi\left({p\over 2} + 1 \pm {1\over 4m}\right) -
\psi\left({p\over 2} \pm {1\over 4m}\right) \right]
%\sum\limits_{k=0}^{\infty}
%{(-1)^k  \over [2m(p+k)+ 2m \pm 1][2m(p+k) \pm 1]}
\label{sol}
\end{equation}
where $\psi(z)$ is the digamma function \cite{abr1}.
This is our main result.
The staircases obtained from formula (\ref{sol})   are
presented in Fig.~\ref{platofg2}. The main staircase
spans the space $0 < \nu < 1/2$ while the secondary staircase
spans the space $0 < \nu < 1/3+\Delta\nu_{1/3} = 0.37187$.
The plateau widths of the most prominent plateaus are presented
in Tables~I-III.

Now we compare with the experiment.
The experimental data show that in 2-dimensional
hole systems  $\Delta\nu_0$ and
$\Delta\nu_{1/3}$ belong to the main sequence.
Indeed $\Delta\nu_0 = 0.2854$ agrees perfectly well with
the experimental value $\sim 0.28$ of the critical filling of the existence
of insulating phase $\sigma_{xy} = 0$ \cite{rod1}.
It is also in perfect agreement with the half-width of the
$\nu = 1$ plateau
$\Delta\nu_1 = \Delta\nu_0 = 0.2854$ \cite{dav1}. In general
the main sequence plateaus (which is the only sequence
present in hole systems) are very well described by the formula (\ref{sol}).

In 2-dimensional electron systems also the secondary sequences
are observed \cite{wil1,saj1,du1}. In fact the experimental data
shows that the range
 $0 < \nu < 1/3+\Delta\nu_{1/3} = 0.37187$ belongs to the secondary staircase
 while the range $ 2/5-\Delta\nu_{2/5} = 0.38127 < \nu
  < 1/2$ belongs to the main staircase.
For example the width of the Wigner
solid region \cite{sha1} is equal to the half-width of the $\nu = 1$ plateau
$\sim 0.19$ which is again very close to our value
$\Delta\nu_0 = 0.1835$ taken from the secondary sequence.
The Hall conductance in the gap between the
 two staircases $0.37187 < \nu < 0.38127$ shows a peculiar behavior
 \cite{rem2}.

For the sake of easy visual comparison with the experimental data
we present the Hall conductivity
$\sigma_{xy}(\nu)$ and resistivity $\rho_{xy}(1/\nu)$ for electron systems
in Fig.~\ref{platofg3} and Fig.~\ref{platofg4}.
We would like to stress that although we can not say which
sequences will be realized for a particular material
we can predict the plateau width {\it if} we know which sequences
are realized.

Further we would like to discuss the conditions
under which the real system data is close to the ideal one described
above. First we discuss the effect of disorder.
In Fig.~\ref{platofg5} we present a schematic dependence
of the half width of the $\nu = 1$ plateau as a function of the
sample mobility $\mu$ at low temperatures and zero magnetic field.
It is a summary of our investigation of the available data.
The mobility is a measure (although indirect) of the disorder
in the sample. At very low mobilities there is no quantum Hall effect.
At mobilities in the range of 1-2$\times  10^{5}$~cm$^{2}$/V.s
the plateau width has a maximum. It is due to the broadening
of the plateaus by disorder-induced local electron reservoirs spread
in the sample. This broadening is reduced when the sample quality is
improved.
Above $\mu$ = 1-2$\times  10^{6}$~cm$^{2}$/V.s  the plateau width
saturates to the value $\Delta\nu_1 \approx 0.19$ ( for
 the best published sample $\mu=1.2\times  10^{7}$~cm$^{2}$/V.s \cite{du1}).
If the disorder would be relevant to the width of the plateaus
one would expect a gradual decrease to zero of $\Delta\nu_1$ when $\mu$
is increased. A measure of the disorder
is the width of the smallest resolvable plateau. For the data
published in Ref.~\cite{du1} it is for example
$\Delta\nu_{5/11} \approx 0.004$. In low density systems the effect
of disorder
is more pronounced. We can estimate it
from the difference between the
experimental width of the Winger insulator region $\approx 0.19$
and our theoretical value 0.1835  to be of order of 0.007.

Next we discuss the effect of temperature. In Fig.~\ref{platofg6}
we present a combined plot of quantum Hall effect data
and Wigner insulator data. On the horizontal axis
the filling factor $\nu$ in the case of Wigner insulator and
and $1-\nu$ in the case of QHE is given. The black points are
the melting temperature $T$ of the Wigner insulator normalized to
the melting temperature of a classical Wigner crystal
$T_{cl} = \sqrt{n_s} e^2/127\varepsilon \sqrt{\pi}$. The data is
from the review of Williams {\it et al.} \cite{sha1} and is obtained
with different techniques in different groups
on samples with densities ranging from 3.4$\times 10^{10}$~cm$^{-2}$
 to 10.2$\times 10^{10}$~cm$^{-2}$. No systematic trend of the
 density dependence of the melting temperature exists.
 On the same plot we give experimental $\sigma_{xy}(1-\nu)h/e^2$
 taken from Sajoto {\it et al.} (solid lines)
 \cite{rem3} and the half width of the $\nu = 1$ plato from
 Clark {\it et al.} \cite{cla1} (circles). The $\sigma_{xy}(1-\nu)h/e^2$
 dependencies has been offset vertically to the corresponding
 normalized temperature. The density of the sample is
 5.5$\times 10^{10}$~cm$^{-2}$
 ($T_{cl} = 380$~mK). The density of the sample of
Clark {\it et al.} is 1.9$\times 10^{11}$~cm$^{-2}$ corresponding
to almost twice higher $T_{cl}$. As it is obvious from the figure
the characteristic temperature scale of the existence of
both IQHE  and Wigner insulator is the melting temperature
of the classical Wigner crystal. Moreover, visual inspection of the
data of Sajoto {\it et al.} and Goldman {\it et al.}
Ref.~\cite{saj1} shows that also the {\it fractional} QHE exists on
the same temperature scale. Typical lowest dilution refrigerator temperatures
are $20-30$~mK $\ll T_{cl}$ which shows that the low temperature limit
is reached experimentally. The connection of the
melting of a Wigner crystal with the width of the plateaus in the integer
QHE was discussed in Ref.~\cite{gro1}.

We would like to mention that here we have neglected the effect of dissipation
($\sigma_{xx} = 0$) and related to it reentrant phases \cite{saj1,san1}.

At the end we would like to discuss the consequences of the
existence of formula (\ref{sol}).\\
1. Integer and fractional Hall effect seem to have the same origin
   as well as the low-filling factor insulating phase (Wigner crystal)
   which can be viewed as $\nu = 0$ quantum Hall effect.\\
2. Disorder is highly improbable to be the reason
    for the finite width of the plateaus.\\
These two conclusions cast serious doubts on the validity
of the standard picture\cite{pra1} of the behavior  of 2DEG in magnetic field.

The symmetry relations $\nu \rightarrow 1\pm \nu$
as well as recursive relations of the type of (\ref{equ})are natural
   for the properties of electrons moving in periodic
    potential \cite{hof1}. It has been shown, however, \cite{tho1}
     that if one uses a
    Kubo formula for the Hall conductivity of independent
 spinless electrons in external periodic potential, one can
 not obtain the fractional QHE.
    The problem we see with the Kubo formula is that it is
    a rewritten fluctuation-dissipation theorem and it is not clear
    if it could be applied for description of a dissipationless
    phenomenon like QHE. For example one can not obtain
 superconductivity by applying the Kubo formula
to a noninteracting electron gas.

I would like to thank G. Sch\"on and A. MacDonald for  stimulating discussions,
H.L. St\"ormer for sending me
preprints, C. Bruder and G. Falci for useful suggestions.
I would like to acknowledge the support of the A. von Humboldt Foundation.
The work is
supported by the Sonderforschungsbereich 195 of the DFG.

    %\noindent

\begin{figure}
\caption{
Schematic drawing of two adjacent QHE plateaus at filling
factors $\nu_a$ and $\nu_b$. The plateaus are symmetrically extended
on both sides of the classical line.
\label{platofg1}}
\end{figure}

\begin{figure}
\caption{
The $\sigma_{xy}(\nu)$ plot of the main sequence staircase $m = 1$
(the doted line)
and the secondary staircase $m = 2$ (the solid line)
obtained using equation
(\protect\ref{sol}) in the region $0 < \nu < 1/2$.
\label{platofg2}}
\end{figure}

\begin{figure}
\caption{
The $\sigma_{xy}(\nu)$ plot of the staircase obtained using equation
(\protect\ref{sol}) and symmetry relations
(\protect\ref{s}) in the region $0 < \nu < 3$.
\label{platofg3}
}
\end{figure}

\begin{figure}
  \caption{
  The $\rho_{xy}(1/\nu)$ plot of the staircase obtained using equation
(\protect\ref{sol}) and symmetry relations
(\protect\ref{s}).
   \label{platofg4}
}
 \end{figure}

\begin{figure}
  \caption{
  Schematic dependence of the plateau half-width $\Delta\nu_1$ on
  the mobility.
   \label{platofg5}
}
 \end{figure}

\begin{figure}
  \caption{
  Combined plot of quantum Hall effect data
and Wigner insulator data. On the horizontal axis is given
the filling factor $\nu$ in the case of Wigner insulator and
and $1-\nu$ in the case of QHE. The black points are
the melting temperature $T$ of the Wigner insulator normalized to
the melting temperature of a classical Wigner crystal
$T_{cl} = \protect\sqrt{n_s} e^2/127\varepsilon
\protect\sqrt{\pi}$. Solid lines
are the  experimental $\sigma_{xy}(1-\nu)h/e^2$
 taken from Sajoto {\protect\it et al.}. Circles
 are the half width of the $\nu = 1$ plato from
 Clark {\protect\it et al.}. The $\sigma_{xy}(1-\nu)h/e^2$
 dependencies has been offset vertically to the corresponding
 normalized temperature.   \label{platofg6}}
 \end{figure}

 \begin{table}
\caption{Plateau half-widths of the main sequence $\nu = p/(2p+1)$.}
\begin{tabular}{c|cccccc}
 $\nu$       & 0      & 1/3    & 2/5    & 3/7    & 4/9    & 5/11 \cr
\tableline
 $\Delta\nu_{\nu}$ & 0.2854 & 0.0479 & 0.0187 & 0.0098 & 0.0060 & 0.0041 \cr
 \end{tabular}
 \end{table}
 \begin{table}
\caption{Plateau half-widths of the secondary sequence $\nu = p/(4p+1)$.}
\begin{tabular}{c|cccc}
 $\nu$       &0      & 1/5    & 2/9    & 3/13   \cr
\tableline
 $\Delta\nu_{\nu}$ &0.1835 & 0.0165 & 0.0057 & 0.0028 \cr
 \end{tabular}
 \end{table}

 \begin{table}
\caption{Plateau half-widths of the secondary sequence $\nu = p/(4p-1)$.}
\begin{tabular}{c|cccc}
 $\nu$       &1/3     & 2/7    & 3/11    & 4/15  \cr
\tableline
 $\Delta\nu_{\nu}$ &0.0385  & 0.0091 & 0.0039  & 0.002 \cr
 \end{tabular}
 \end{table}


\begin{references}
\bibitem{rem1} Other factors are the measuring current and
the shape of the confining potential. The measuring current can
be made low enough so that the plateaus' width does not change any more.
The shape of the confining potential depends on the particular
structure. We will consider here single heterojunction systems.
\end{references}
     \end{document}